\documentclass[aps,letterpaper,twocolumn,prl,footinbib,floatfix]{revtex4}
\usepackage{amsmath}
\usepackage{amsfonts}
\usepackage{amssymb}
\usepackage[scanall]{psfrag}
\usepackage{psfrag}
\usepackage{graphicx}
\usepackage{color}
\usepackage{tabularx}

\begin{document}
\newcommand{\of}[1]{\left( #1 \right)}
\newcommand{\sqof}[1]{\left[ #1 \right]}
\newcommand{\abs}[1]{\left| #1 \right|}
\newcommand{\avg}[1]{\left< #1 \right>}
\newcommand{\cuof}[1]{\left \{ #1 \right \} }
\newcommand{\bra}[1]{\left < #1 \right | }
\newcommand{\ket}[1]{\left | #1 \right > }
\newcommand{\pil}{\frac{\pi}{L}}
\newcommand{\bx}{\mathbf{x}}
\newcommand{\by}{\mathbf{y}}
\newcommand{\bk}{\mathbf{k}}
\newcommand{\bp}{\mathbf{p}}
\newcommand{\bl}{\mathbf{l}}
\newcommand{\bq}{\mathbf{q}}
\newcommand{\bs}{\mathbf{s}}
\newcommand{\psibar}{\overline{\psi}}
\newcommand{\svec}{\overrightarrow{\sigma}}
\newcommand{\dvec}{\overrightarrow{\partial}}
\newcommand{\bA}{\mathbf{A}}
\newcommand{\bdelta}{\mathbf{\delta}}
\newcommand{\bK}{\mathbf{K}}
\newcommand{\bQ}{\mathbf{Q}}
\newcommand{\bG}{\mathbf{G}}
\newcommand{\bw}{\mathbf{w}}
\newcommand{\bL}{\mathbf{L}}
\newcommand{\ohat}{\widehat{O}}
\newcommand{\up}{\uparrow}
\newcommand{\down}{\downarrow}
\newcommand{\MM}{\mathcal{M}}
\newcommand{\tX}{\tilde{X}}
\newcommand{\tY}{\tilde{Y}}
\newcommand{\tZ}{\tilde{Z}}
\newcommand{\tOm}{\tilde{\Omega}}
\newcommand{\barA}{\bar{\alpha}}

\author{Eliot Kapit}
\email{ekapit@mines.edu}
\affiliation{Department of Physics, Colorado School of Mines, Golden, CO}
\affiliation{Quantum Engineering Program, Colorado School of Mines, Golden, CO}

\author{Vadim Oganesyan}
\affiliation{Center for Computational Quantum Physics, Flatiron Institute, 162 5th Avenue, New York, NY 10010, USA}
\affiliation{Physics program and Initiative for the Theoretical Sciences,
The Graduate Center, CUNY, New York, NY 10016, USA}
\affiliation{Department of Physics and Astronomy, College of Staten Island, CUNY, Staten Island, NY 10314, USA}

\title{Small logical qubit architecture based on strong interactions and many-body dynamical decoupling}

\begin{abstract}
We propose a novel superconducting logical qubit architecture, called the Cold Echo Qubit (CEQ), which is capable of preserving quantum information for much longer timescales than any of its component parts. The CEQ operates fully autonomously, requiring no measurement or feedback, and is compatible with relatively strong interaction elements, allowing for fast, high fidelity logical gates between multiple CEQ's. Its quantum state is protected by a combination of strong interactions and microwave driving, which implements a form of many-body dynamical decoupling to suppress phase noise. Estimates based on careful theoretical analysis and numerical simulations predict improvements in lifetimes and gate fidelities by an order of magnitude or more compared to the current state of the art, assuming no improvements in base coherence. Here, we consider the simplest possible implementation of the CEQ, using a pair of fluxonium qubits shunted through a shared mutual inductance. While not necessarily the best possible implementation, it is the easiest to test experimentally and should display coherence well past breakeven (as compared to the limiting coherence times of its components). A more complex three-node circuit is also presented; it is expected to roughly double the coherence of its two-fluxonium counterpart.
\end{abstract}

\maketitle

\section{Introduction and circuit design}

While superconducting quantum computer architectures \cite{krantz2019quantum} have made enormous progress over the last few years, performance is generally still limited by the short coherence times of individual qubits. To address this, researchers have explored a range of more complex small logical qubit designs, which include both passive, topologically protected circuits \cite{PRXQuantum.2.010339,aghaee2022inas} and active error correction codes based on engineered dissipation or measurement and feedback \cite{ofekpetrenko2016,kapit2016,hu2019quantum,cai2021bosonic,ni2022beating,sivak2022real}. Alongside these approaches are continuously driven schemes such as the two-leg cat qubit, which can \emph{exponentially} suppress error along one axis at the cost of linearly increasing it along another \cite{lescanne2020exponential,berdou2022one}, requiring dissipation but no active measurement. However, these approaches all present serious challenges, including exotic circuit elements, physically large 3d devices, complex measurement and control sequences, and high symmetry requirements in fabrication. While surmountable, these obstacles make scaling these designs to large quantum computer systems difficult.

In this letter, we propose a new approach: a small logical qubit design based on strong interactions and continuous driving, which can suppress \emph{all} single qubit error channels fully autonomously, with no dissipative elements. We call our design the Cold Echo Qubit (CEQ). It consists of just two or three flux-type qubits (such as capacitively shunted flux qubit \cite{yangustavsson2015} or fluxonium \cite{manucharyan2009fluxonium,xiong2022arbitrary,somoroff2021millisecond,zhang2021universal} devices), coupled by strong, fixed longitudinal interactions and continuously driven through a strong charge drive. We will show that this circuit is robust to small variations in the device parameters, has a single tone control structure, and is compatible with tunable couplers, so that fast gates can be enacted.

\begin{figure}
\includegraphics[width=2in]{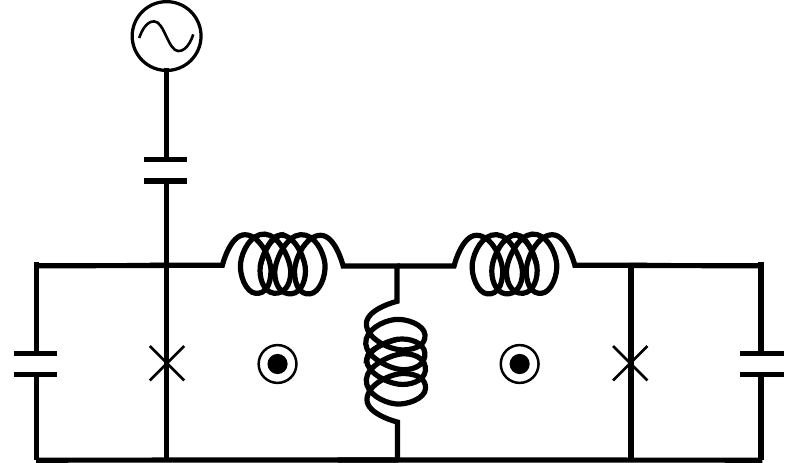}
\includegraphics[width=1.0in]{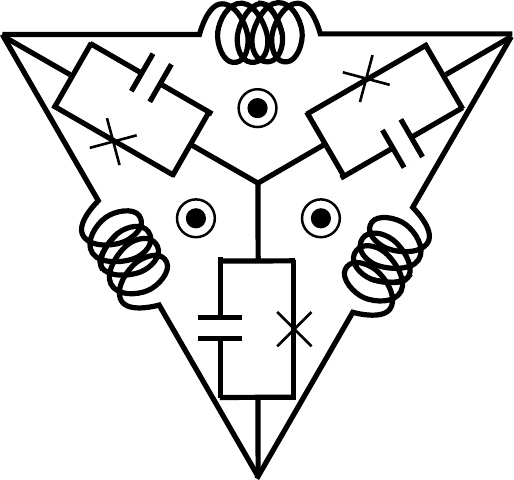}
\caption{Circuit diagrams for the logical qubits considered in this work. Left: two-node circuit inspired by \cite{kou2017fluxonium}; the central shunt inductance is reduced by a factor of $\gamma <1$ compared to the inductances in each component fluxonium. Equal and opposite flux biases break the degeneracy of the antiferromagnetic doublet ground state. Right: a three-node version, where the inductors create both the inductive potential for each fluxonium and maximize the (now ferromagnetic) couplings to the other fluxonia.}\label{circuitfig}.
\end{figure}

Inspired by the fluxonium molecule circuit demonstrated in \cite{kou2017fluxonium}, we first analyze a minimal version of this device, with the same basic circuit topology, operated as a logical qubit via asymmetric flux biases and continuous driving. The circuit, shown in FIG~\ref{circuitfig}, consists of a pair of fluxonium qubits; the base qubits' inductances are shunted to ground through a common inductor $\gamma E_L$, with $\gamma \leq 1$. Using standard quantization methods, we arrive at phase basis Hamiltonian:
\begin{eqnarray}
H &=& - 4 E_C \of{\frac{\partial^2}{\partial \phi_1^2} + \frac{\partial^2}{\partial \phi_2^2}   } \\
& & - E_J \of{\cos \of{\phi_1 + \Phi_1} + \cos \of{\phi_2 + \Phi_2}   } \nonumber \\
& & + E_L \frac{1+\gamma}{2 \of{2+\gamma}} \of{\phi_1^2 + \phi_2^2} + \frac{E_L}{2+\gamma} \phi_1 \phi_2. \nonumber
\end{eqnarray}
Here, $\Phi_1$ and $\Phi_2$ are external control fluxes, both of which are to be biased near $\pi$ for CEQ operation, $\phi_1$ and $\phi_2$ are the quantum phase variables and we assume $E_J \gg E_L$ and $E_J > E_C$.  While we assume these parameters are symmetric for simplicity, inevitable small variations will not significantly affect logical qubit performance.

The CEQ is a strongly coupled device, and with two or three degrees of freedom it can be straightforwardly diagonalized numerically. However, it is much more illuminating--particularly for analyzing the noise model and choosing device parameters--to diagonalize each qubit separately to construct a reduced, two-spin Hamiltonian. In the limit $E_J \gg E_L$, with external flux $\Phi = \pi$ the fluxonium Hamiltonian consists of a double-well potential with minima $\phi_m \simeq \pm \of{\pi - E_{Ls}/E_J}$, where $E_{Ls} \equiv E_L \frac{1+\gamma}{2 \of{2+\gamma}} $ is the single fluxonium inductive energy. If we move away from the symmetry point by letting $\Phi = \pi + \delta \Phi$ (for small $\delta \Phi$), the two minima locations shift by $ \delta \Phi$, with corresponding energy shifts $\simeq \pm \pi E_{Ls} \delta \Phi$. If we let the persistent current states near $\pm \pi$ be $\ket{0}$ and $\ket{1}$, we can represent this bias term as $h \of{\Phi} \sigma^z$. Given an inter-well tunneling term $- \kappa \sigma^x$, we have $H_0 = - \kappa \sigma^x + h \of{\Phi} \sigma^z$. Here $\kappa$ can be computed in the WKB approximation and decreases exponentially in $\sqrt{E_J/E_C}$; at the symmetry point the single qubit excitation energy is $2\kappa$. Quantizing the system in a two-qubit basis, we obtain
\begin{eqnarray}\label{baseH}
\frac{E_L}{2+\gamma} \phi_1 \phi_2 \simeq \frac{E_L}{2+\gamma} \of{\pi - E_{Ls}/E_J}^2 \sigma_1^z \sigma_2^z \equiv J \sigma_1^z \sigma_2^z. \\
H_0 = - \kappa \of{ \sigma_1^x+\sigma_2^x} + h \of{\Phi_1} \sigma_1^z + h \of{\Phi_2} \sigma_2^z + J \sigma_1^z \sigma_2^z.
\end{eqnarray}
Higher levels are generally far enough away in energy to be ignorable here. The spectrum of $H_0$ consists of the symmetric and antisymmetric combinations of $\ket{01}$ and $\ket{10}$, with energies $-J \pm \Omega_0$, separated from the equivalent combinations of $\ket{00}$ and $\ket{11}$ with energies $+J \pm \Omega_0$, where $\Omega_0 \simeq \kappa^2/J$. We let our logical states be the perturbatively dressed antiferromagnetic doublet, $\ket{0_L} \simeq \ket{01}$ and $\ket{1_L} \simeq \ket{10}$, and choose $h \of{\Phi_1} = - h/4$ and $h \of{\Phi_2} = +h/4$, detuning the logical states by $h$ so that they are nearly $z$ eigenstates. We assume $h \gg \Omega_0$. Note this is a significant difference from \cite{kou2017fluxonium}, where there only a single, global flux control was used. The $L=3$ design, detailed in the supplemental material, uses exactly the same principles and drive structure, though the interactions are ferromagnetic in that case.

To motivate this design, we now introduce noise. We use a noise model drawn from the detailed  spectroscopy reported in \cite{bylandergustavsson2011,yangustavsson2013,yangustavsson2015,quintana2017observation,dai2022dissipative} for flux qubits, which are phenomenologically similar to fluxonium devices. Every component qubit couples a thermal bath though a mix of $x$, $y$, and $z$ couplings. For flux-like qubits the $x$ coupling is generally neglible, and the $z$ coupling scales as the qubit's persistent current (see \cite{yangustavsson2015} for a survey); consequently for fluxonia $y$ error (the charge channel) is largest. These are entangling interactions with a thermal environment at temperature $T_{eff}$; this temperature is often larger than the nominal fridge temperature so $T_{eff}$ is the temperature seen by the qubits themselves.

In this DC form, the CEQ is well-protected against single spin flip excitations from the environment, since they cost $\sim 2J \gg k_B T_{eff}$ (where $T_{eff}$ is the effective bath temperature seen by the qubits) and are thus very strongly suppressed. It is also protected against low-energy transitions between the two logical states $\ket{0_L}$ and $\ket{1_L}$ by virtue of small perturbative matrix elements for $\bra{1_L} \sigma_j^y \ket{0_L}$ and $\bra{1_L} \sigma_j^z \ket{0_L}$. 

However, it is \emph{not} protected at all against low-frequency flux noise dephasing. Alongside these bath couplings, each qubit experiences independent, classical $1/f$ noise along $z$, represented by fluctuating terms $\delta h_j \of{t} \sigma_j^z$ in the Hamiltonian. In flux qubits, this flux noise has been extensively studied, and is very well-described by $1/f^\alpha$ (with $\alpha \simeq 1$) until $f \sim 1$GHz, where it crosses over to ohmic-spectrum bath interactions \cite{yangustavsson2015,quintana2017observation,dai2022dissipative}. To remedy this, we resonantly Rabi drive \cite{martinisnam2003} between $\ket{0_L}$ and $\ket{1_L}$, echoing out this final remaining error channel. As target Rabi frequencies for this process are in the tens of MHz, we will remain firmly in the classical regime for this channel.

In principle, the AC variation of nearly any term can drive transitions between $\ket{0_L}$ and $\ket{1_L}$, including oscillating $Z$ biases, or the \emph{magnitude} of one or more transverse fields by replacing the single junctions in FIG.~\ref{circuitfig} with SQUIDs. However, the resulting flux crosstalk can introduce significant calibration challenges. To simplify things, we achieve similar results using single junctions and AC charge driving. We will now show that, for a given $\Omega_0$, the maximum achievable $\Omega_{AC} \simeq 0.6 \Omega_0$ is sufficient to produce long lifetimes given base coherences reported in the literature.

To derive this limit, we consider the Hamiltonian of a single fluxonium qubit, with an external charge offset $Q_{ext}$ across the junction. The Hamiltonian becomes
\begin{eqnarray}\label{Hsingle}
H =  4 E_C \of{-i \frac{\partial}{\partial \phi} + \frac{Q_{ext}}{2e} }^2 - E_J \cos \of{\phi + \Phi} + \frac{E_{Ls}}{2} \phi^2
\end{eqnarray}
Now, a \emph{static} $Q_{ext}$ has no physical consequences, as it can be eliminated by a simple unitary transformation. If we apply $\ket{\psi \of{\phi}} \to e^{-i  \frac{Q_{ext}}{2e} \phi } \ket{\psi \of{\phi}}$ then we can remove $Q_{ext}$ from Eq.~\ref{Hsingle} entirely;\footnote{This is only possible because the quadratic inductive potential has broken the periodicity of $\phi$; in superconducting qubits where this is not the case, this periodicity requirement makes such a unitary transformation impossible to construct and the qubit energies do become sensitive to offset charge as a result.} viewing Eq.~\ref{Hsingle} as a single particle in a double well potential, the charge offset acts as constant vector potential $\mathbf{A} = \frac{Q_{ext}}{2e} \hat{\phi}$. However, if $Q_{ext}$ varies in time, it cannot be eliminated in the time-dependent Schrodinger equation and yields nontrivial dynamics. 

Identifying the two persistent current states as our $z$ basis, quantum tunneling between the wells creates a transverse field $x$ term. And while a charge offset does not change the \emph{magnitude} of this process, the ``particle" does accumulate a complex Peierls phase $\varphi$ as it tunnels between the two wells, where $\varphi \simeq \int \mathbf{A} \cdot d \phi \simeq 2 \phi_m Q_{ext}/2e$. This rotates our $x$ field about $z$:
\begin{eqnarray}\label{H0rot}
H_0 \to \kappa \of{\cos \of{\varphi \of{t}} \sigma^x + \sin \of{\varphi \of{t}} \sigma^y} + h \of{\Phi} \sigma^z.
\end{eqnarray}
In other words, charge driving causes the transverse field to wobble back and forth in the $x-y$ plane. As shown in the supplemental material, this relationship holds even if $Q_{ext}$ is varied at frequencies that far exceed $\kappa$.

This is one way to implement the Hamiltonian in RFQA-D \cite{kapit2021noise}, the quantum optimization scheme that inspired the CEQ. Driving just one of the two qubits\footnote{In this simple setup, there is no benefit to driving both and the maximum achievable Rabi frequency $\Omega_{AC}$ does not appear to be increased by doing so. There are more exotic ways of engineering or operating this design that can overcome this limit, such as adding SQUIDs to the circuit and/or driving at very high frequencies, but those are beyond the scope of this writeup.} and the transverse field rotates in the $x-y$ plane with angle $\varphi \of{t}$; noting that $\Omega_0 \simeq \kappa^2/J$ the low-energy effective Hamiltonian is given by
\begin{eqnarray}
H \of{t} \simeq \Omega_0 \sigma_1^x \of{\cos \varphi \of{t} \sigma_2^x + \sin \varphi \of{t} \sigma_2^y} + \frac{h}{4} \of{ \sigma_1^z - \sigma_2^z}
\end{eqnarray}
As shown in \cite{kapit2021noise}, if we let the angle $\varphi \of{t} = \barA \sin \of{2\pi h t}$, this produces a resonant drive term between the two logical states with an amplitude given by $\frac{\Omega_0}{2} f \of{\barA}$. $f \of{\barA}$ is a Bessel function that peaks at $\simeq 1.18$ around $\barA \simeq 1.84$, leading to a maximum Rabi rate $\Omega_{AC} \sim 0.6 \Omega_0$. The amplitude of the charge drive here is on the order of half a Cooper pair of offset charge across the junction; driving at higher amplitudes is counter-productive, decreasing $\Omega_{AC}$ while potentially introducing new error sources. This drive term strongly suppresses low-frequency phase noise, sampling it at $\sim 2 \Omega_{AC}$, mitigating the last error channel for a high coherence logical qubit.

\begin{figure*}
\includegraphics[width=3in]{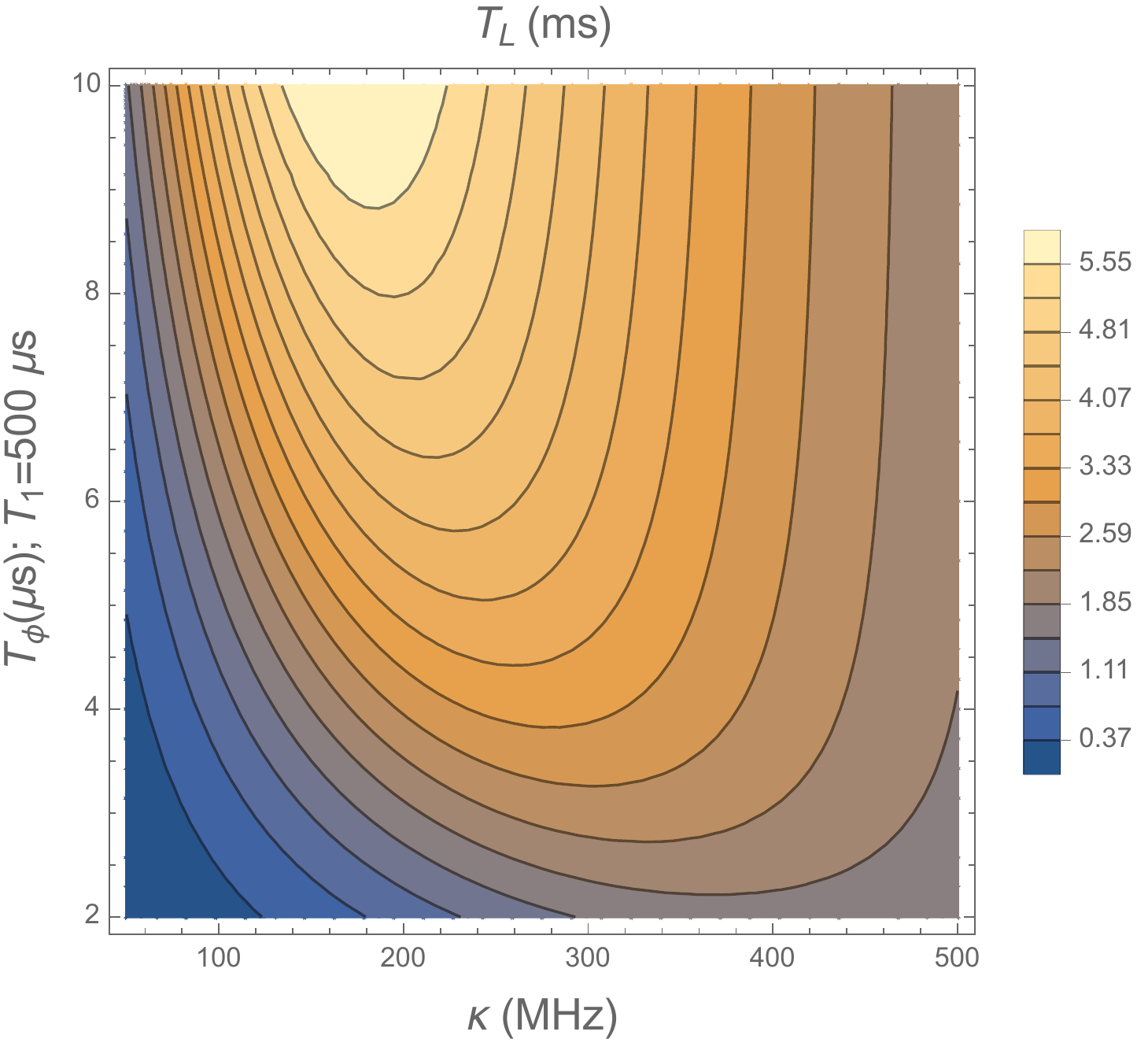}
\includegraphics[width=3in]{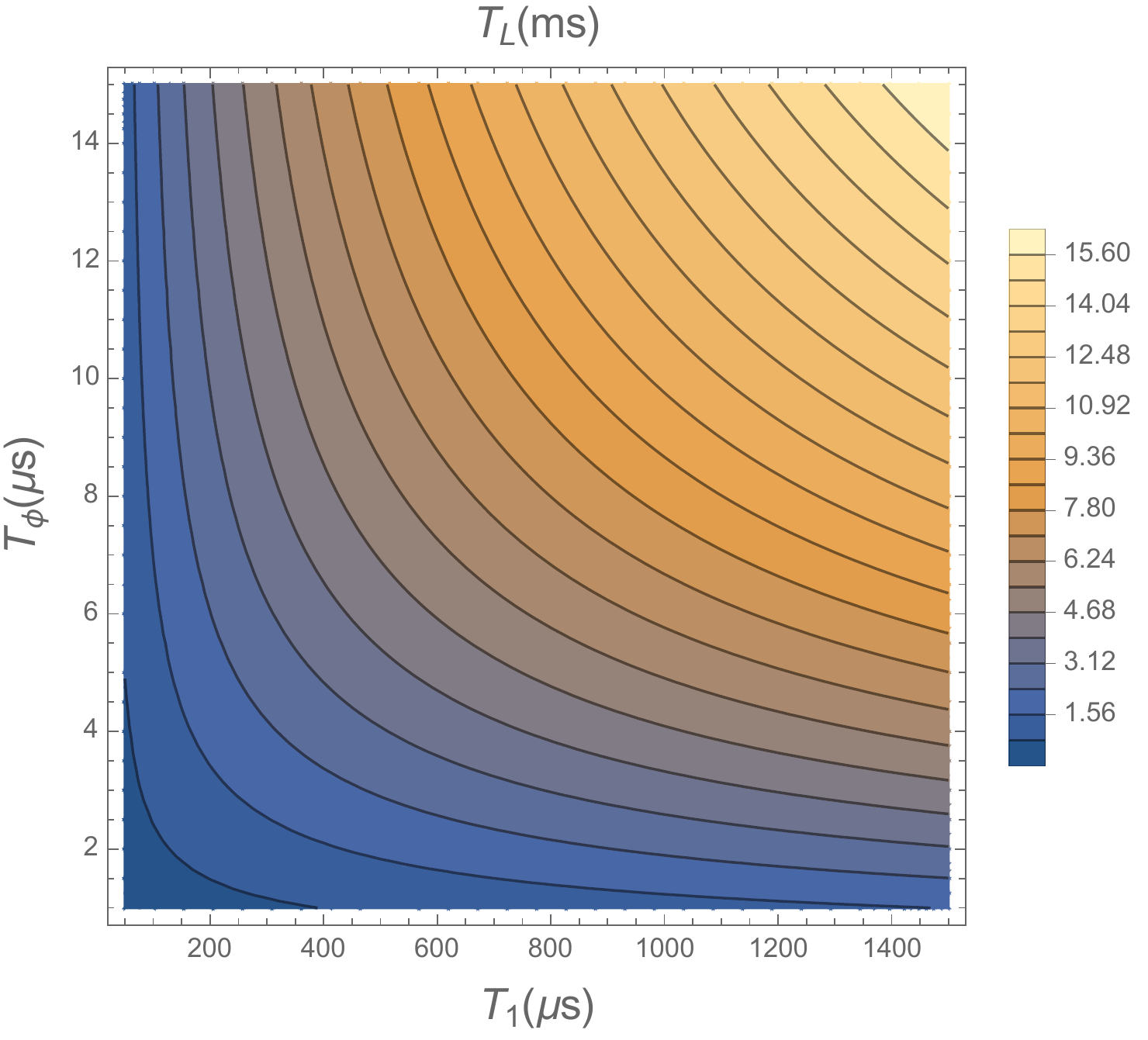}
\caption{Logical state lifetimes, for an $L=2$ CEQ with $J= 2\pi \times 1.5$ GHz and $k_B T_{eff} = 2\pi \times 0.5$ GHz, assuming $\Omega_{AC} = 0.6 \Omega_{0}$ as described in the text. On the left, logical lifetimes as a function of transverse field $\kappa$ and off-sweet-spot $T_\phi$, for single-qubit $T_1 = 500 \mu$s, showing that $T_L$ does peak at finite $\kappa$ for a given $T_1$ and $T_\phi$, but not sharply so. On the right, best achievable lifetime (varying $\kappa$) as a function of $T_1$ and $T_\phi$. For simplicity, the energy-dependence of $T_1$ is ignored in low-energy relaxation and, appropriate to fluxonia, $\Gamma_z$ is considered to be negligible (it can be further suppressed by increasing the detuning $h$, which does not enter into the other error rates). For simplicity, $T_L$ is calculated as the inverse of the sum of all error rates in Eqs.~(\ref{thermerr}-~\ref{oneferr}); note that error is substantially anisotropic in some regions of these plots. As shown in the supplemental material, for $L=3$, due to reduced high and low energy error through the charge channel, lifetimes at the higher end of the base coherence range can be more than double those of the $L=2$ counterpart.}\label{TLfig}
\end{figure*}

\section{Logical Coherence estimates}

To prove these claims, we now consider each noise channel in detail, for $L=2$ and 3. Let the nominal relaxation rate through $y/z$ to the environment for a \emph{single component qubit} be $\Gamma_{y/z}^s$ (measured at the flux symmetry point). We begin with \textbf{Thermal excitation error}, where energy is absorbed from the environment to create single-flip excitations. This is predominantly through the $y$ channel. As a single excitation can scramble the logical state, our thermal error rate $\Gamma_{th}$ is given by
\begin{eqnarray}\label{thermerr}
\Gamma_{th} \simeq L \Gamma_y^s  \frac{ c_L J}{2 \kappa } \exp \of{- \frac{c_L J}{k_B T_{eff}}},
\end{eqnarray}
where $c_L=2$ for $L=2$ and $4$ for larger $L$. The factor of $c_L J/\kappa$ comes from assuming ohmic charge noise and that the qubit's $T_1$ is measured at energy $2\kappa$ for single fluxonia at the symmetry point. This error rate thus decreases exponentially with linear increases in $J$.

\textbf{Low energy environmental error} arises from higher order perturbative corrections, as the matrix elements $\bra{1_L} \sigma_j^{x/y/z} \ket{0_L}$ are all generically nonzero. A simple perturbative analysis in the DC limit yields
\begin{eqnarray}\label{leerr}
\Gamma_{z,le} \simeq  4 L \Gamma_z^s \of{ \frac{\Omega_0}{h}}^2, \; \; \Gamma_{y,le} \simeq  L \Gamma_{y}^s \of{\frac{\Omega_0}{\kappa}}^2.
\end{eqnarray}
Note that $\Omega_0$ enters this equation quadratically, the result of the perturbative matrix element being $\Omega_0/h$ and bath error rates being derived from Fermi's Golden Rule (where the interaction is squared). In AC operation, as derived in the supplemental material, there can be small shifts to these rates but the overall scaling is identical. To verify these claims, we have performed 
numerical simulations of an $L=2$ CEQ coupled to a broad-spectrum bath; see the supplemental material for details.


$1/f$-\textbf{induced dephasing} is suppressed by the AC drives. We let the \emph{single qubit} pure dephasing time from $1/f$ noise be $T_\phi$, measured via Ramsey decay assuming $T_1 \to \infty$. Rabi driving converts the Gaussian decay of $1/f$ dephasing to simple exponential, and if the Rabi rate $\Omega$ is larger than $1/T_{\phi}$ the resulting $T_{2} \propto T_{\phi}^2 \Omega$. We sum the $1/f$ traces of each qubit to produce a single fluctuating $\delta h \of{t}$, which also has a $1/f$ spectrum; then
\begin{eqnarray}\label{oneferr}
\Gamma_{z,1/f} \simeq \frac{L}{5 T_{\phi}^2 \Omega_{AC} }.
\end{eqnarray}
The factor of $5$ in the denominator comes from numerical best fitting. Note that the $L=2$ CEQ is insensitive to ``common mode" flux noise (uniform for both qubits) \cite{kou2017fluxonium}.

Finally, \textbf{Readout error} is the error in measuring the full CEQ logical state. In the simplest implementations, this is done dispersively by capacitive coupling to a readout resonator \cite{PhysRevLett.95.060501}, with the AC charge drive turned off. As readout error is primarily limited by $T_{1}$ loss during the readout process itself, the increased effective $T_1$ (which does not depend on AC driving at all) should proportionally reduce readout error.

In FIG.~\ref{TLfig}, we plot the logical state lifetime for realistic $T_1$ and $T_\phi$ ranging from ``average" to a bit below the current state of the art. In all cases it is very easy for the CEQ to exceed the current state of the art for single fluxonium devices. This is intrinsically a simplified treatment, and ignores more complex and subtle relationships such as the dependence of $T_{\phi}$ on $J$ ($T_{\phi}$ decreases with increasing $J$, since it decreases with increasing $E_L$, but can vary widely from one design to the next); nonetheless, these plots give reasonable performance expectations for this device. For example, using base coherences $\cuof{T_1,T_\phi} = \cuof{315,4} \mu {\rm s}$ from \cite{zhang2021universal}, and the parameters in FIG.~\ref{TLfig} we obtain a maximum $T_L \simeq 2.5$ms for $L=2$ and $\sim 5$ ms for $L=3$. Interestingly, since $\Gamma_y \gg \Gamma_z$ in fluxonia, at the ``optimal" $T_L$ the CEQ has relatively isotropic \emph{logical} error; in contrast, error in a version based on another flux qubit type (with larger persistent current, and thus, $\Gamma_z$) can be strongly biased along $z$.

\section{Logical gates and outlook}

We now consider logical operations. The base gate set of CEQ 
consists of single qubit logical $x$ and $z$ rotations, along with CPHASE/CZ. The drive terms that generate $\Omega_{AC}$ causes the logical qubit state to continuously rotate around the $x$ or $y$ axis (depending on convention; we choose $x$ in this section), at rates rapid enough to produce very high fidelity gates. So $x$ rotations are performed by simply letting the system evolve under the drive for appropriate amounts of time. Logical $z$ rotations can be performed by pulsing the individual $z$ biases. It may be desirable to momentarily weaken or turn off the $\Omega_{AC}$ field generating continuous $x$ rotations during this step. Note that this does \emph{not} interfere with the $1/f$ protection if the $z$ rotations are short (we expect 10-20 ns), since the duration will be so short that any $1/f$ noise will be sampled at very high frequency as a result, where it is weak.

CZ can be implemented through flux-tunable mutual inductances between pairs of logical qubits. As in logical $z$ rotations, we can either momentarily weaken or disable $\Omega_{AC}$ during this step, or oscillate the coupler at the appropriate frequencies. Unlike transmons, assuming crosstalk is mitigated, a single CEQ can participate in multiple CZ gates simultaneously. This is a potentially significant benefit when incorporating CEQs into a surface code \cite{fowlersurface}; it reduces the length of the error detection cycle and makes the decoding process easier. Note however that in fluxonium qubits, the persistent current is significantly reduced compared to traditional flux qubits, making the strengths of such mutual inductances much weaker, though even a few MHz is sufficient for a coherence limited error rate below $10^{-4}$. Gates of this type are also trivially error transparent \cite{kapit2018error} and error divisible \cite{perez2021error}. A variable \emph{capacitance} between two CEQs could also be implemented (for example by using a flux tunable qubit to mediate interactions as in \cite{sung2021realization}), though the analysis is more complex in that case.

In conclusion, we have proposed a strongly coupled and strongly driven circuit capable of achieving significant coherence improvements against the empirical flux qubit noise model. It requires no new circuit elements and could be tested immediately. The original idea for this design arose from a careful first-principles analysis of noise in AC-driven quantum annealing--hardly a system known for high coherence. We suggest that while the combination of strong coupling \emph{and} strong AC drives makes noise analysis considerably more difficult than the weak limit normally considered, if these results are borne out in experiments there are likely many more surprises to be found in this regime.

\section{Acknowledgements}

We would like to thank Michel Devoret, Steven Dissler, Paul Varosy, David Schuster and Steven Weber for useful discussions. EK's research in this area was funded by NSF grant PHY-1653820 and by ARO Grant No. W911NF-18-1-0125. EK and VO were jointly supported by DARPA under the Reversible Quantum Machine Learning and Simulations program, contract HR00112190068. The Flatiron Institute is a division of the Simons Foundation.

\bibliography{fullbib}

\newpage

\section{Supplemental information}

\subsection{A higher coherence three-variable circuit}

We can achieve even greater coherence in a three-node circuit, at the cost of additional complexity. The optimal circuit, in terms of compactness and maximizing interaction energies for a given $E_L$, is shown on the right in FIG.~\ref{circuitfig}. We assume all the junctions, inductances, and capacitances are equal. We let the phase variables $\phi_j$ be the phase across each Josephson junction (relative the central point, which we let be ground), and define external fluxes $\Phi_j$ (which will all be set at or near $\pi$). In this notation, the circuit Hamiltonian can be written as
\begin{eqnarray}\label{H3}
H &=&  4 E_C \sum_j \of{-i \frac{\partial}{\partial \phi_j} + \frac{Q_{j,ext}}{2e} }^2 \\
& & - E_J \sum_j \cos \of{\phi_j - \Phi_j}  + E_L \sum_j \of{\phi_j^2 - \phi_j \phi_{j+1}}. \nonumber
\end{eqnarray}
From this we derive a single-fluxonium inductance $E_{Ls} = 2 E_L$ and, in the spin basis, an interaction energy $J=E_L \of{\pi - E_{Ls}/E_J}^2$. Note that, due to the different geometry of the circuit elements, the interaction is ferromagnetic in this case.

The design and operation of the $L=3$ CEQ mirrors that of the simpler $L=2$ version, with a few important changes. The local flip energy cost, which sets the exponential suppression of high energy errors, is doubled compared to the $L=2$ circuit. The various logical error rates all experience a 50\% increase in their prefactor due to there being three component qubits that can suffer errors. However, for fixed $\Delta_{local} = 2J/4J$, single qubit $T_{\phi}$ is expected to \emph{increase} compared to the $L=2$ circuit, because a smaller $E_L$ is required to achieve it. Similarly, since a substantially larger value of $\kappa$ is required to achieve a given value of $\Omega_0$, for fixed $\Omega_0$ the low-energy error rate from $y$ errors is reduced, as it is suppressed by $\of{\Omega_0/\kappa}^2$. For resonant charge driving, the limit $\Omega_{AC} \leq 0.6 \Omega_0$ is unchanged, and $\kappa$ cannot exceed $J$ as at that point there is a transition to a paramagnetic ground state and the arguments here no longer hold. The result of all these considerations is increased logical state coherence, as shown in FIG.~\ref{TLfigL3}.

\begin{figure}
\includegraphics[width=3in]{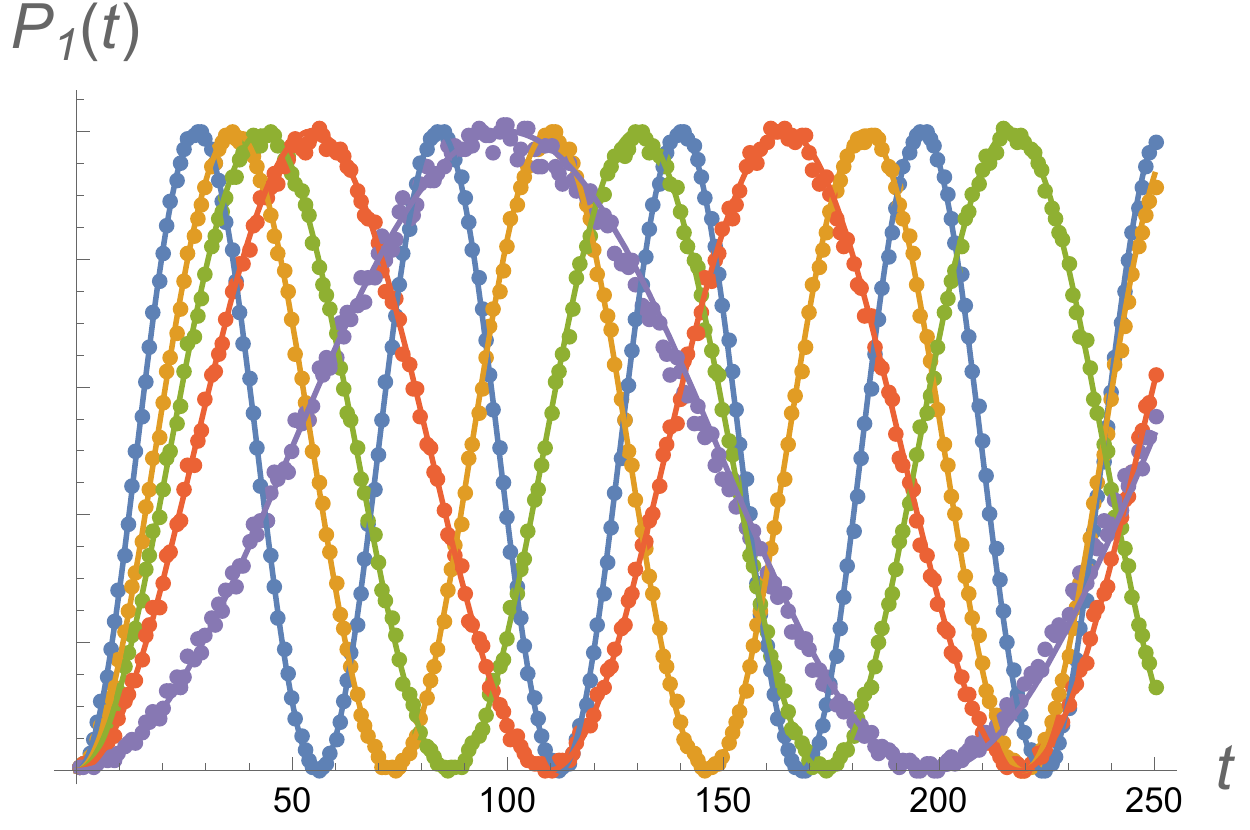}
\caption{Rabi flopping under a very high frequency charge drive. As described in the text, this numerical experiment confirms our predictions for transverse field rotation even at very high frequencies. The blue curve corresponds to the undriven problem, where the system is initialized in $\ket{0_x}+\ket{1_x}$ and the population of $\ket{1_z}$ is tracked. Gold, green, red, and purple data correspond to a very high frequency drive with $\barA = \cuof{1,1.25,1.5,1.9}$, which suppresses the DC average value of the transverse field. Solid lines correspond to the prediction in Eq.~\ref{Htavg} (with no free parameters), showing excellent agreement.  }\label{chargdrive}
\end{figure}

\begin{figure*}
\includegraphics[width=3in]{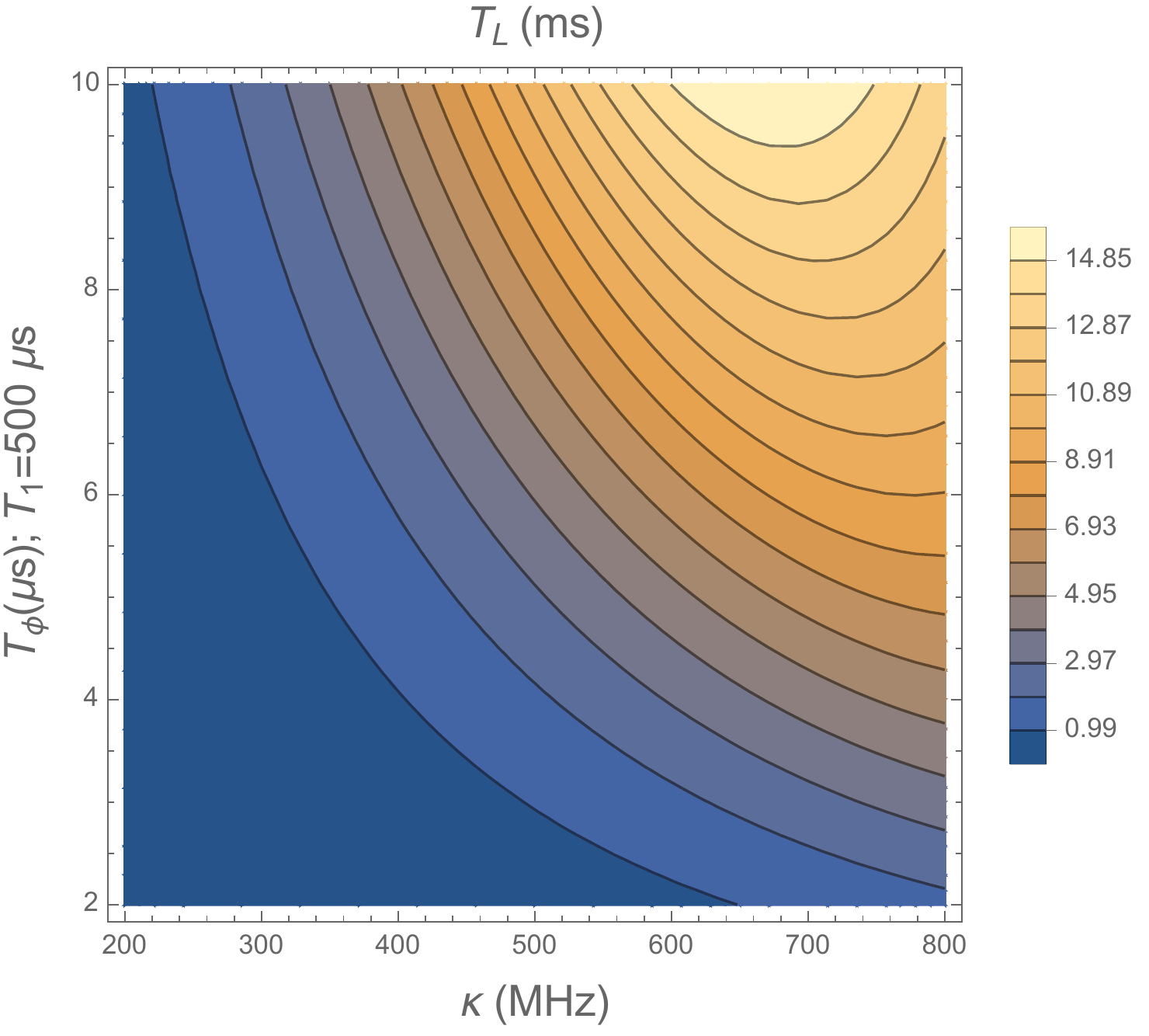}
\includegraphics[width=3in]{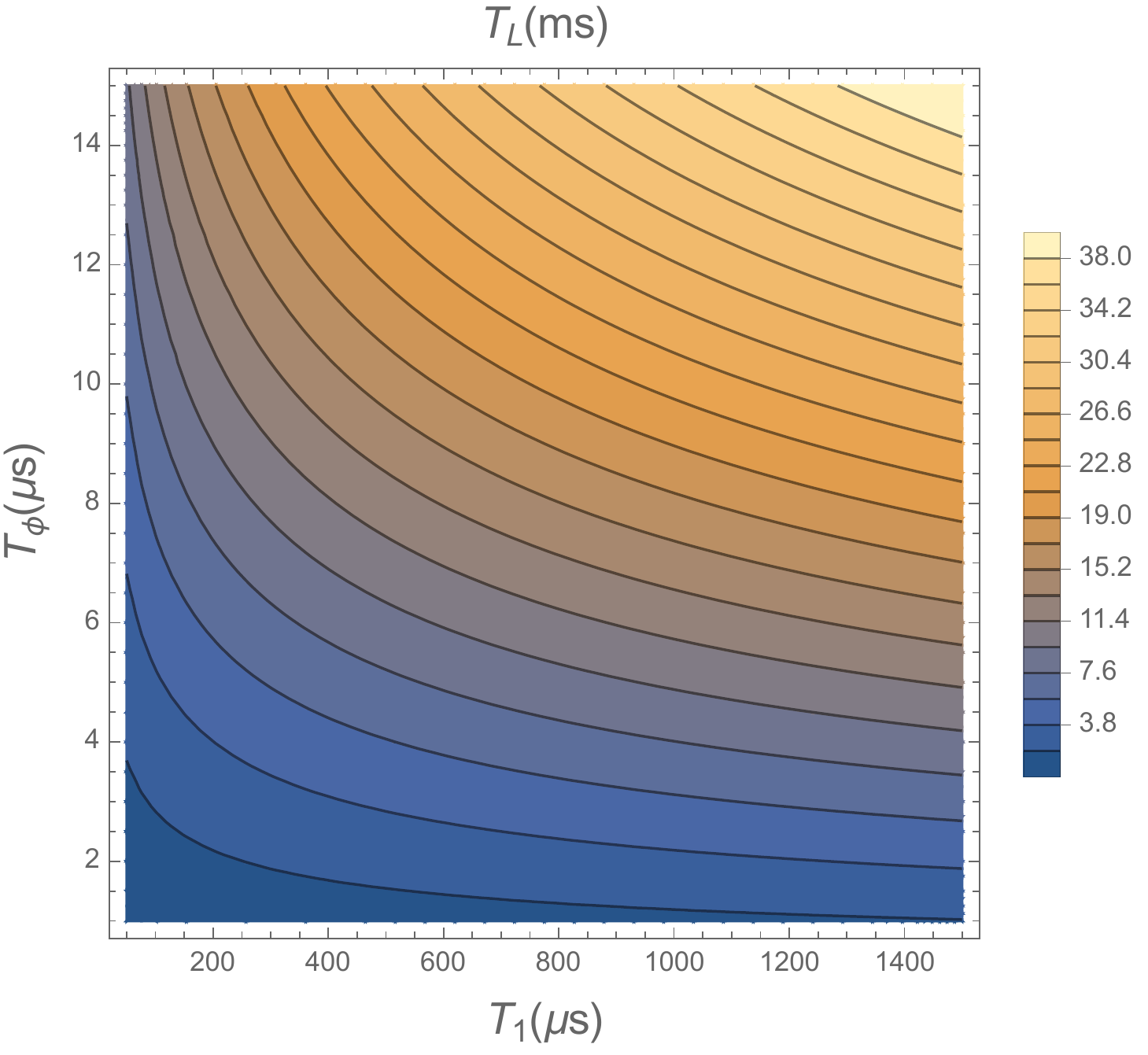}
\caption{Logical state coherence for $L=3$, with all parameters otherwise identical to those of FIG.~\ref{TLfig}.}\label{TLfigL3}
\end{figure*}

\subsection{More details on charge driving}

In the main text, we suppressed $1/f$ noise with a strong AC charge drive, which caused the transverse field to tip back and forth in the $x-y$ plane. However, one might be concerned that other subtle effects break this relationship when the charge offset varies rapidly, and in this section, we provide numerical evidence to show that our arguments hold for frequencies far beyond what one would employ in the CEQ. We can check the fast rotation of the transverse field direction in a highly nontrivial manner by considering \emph{very high frequency} driving. Specifically, let us consider a single fluxonium at the symmetry point (so that the qubit energy is $2 \kappa$ and $H=-\kappa \sigma^x$), and let $E_2$ be the energy of the second excited state. We now apply a charge drive at frequency $\omega$, where $\kappa \ll \omega$ but $\omega \ll E_2$. Such a drive is far off resonant with any transitions, and so 

To verify that the transverse field \emph{direction} follows the charge offset (as specified in Eq.~\ref{H0rot}) in real time dynamics, and to set expectations for the noise analysis below, we performed the following simulation. We numerically diagonalized the single fluxonium Hamiltonian Eq.~\ref{Hsingle} in the phase basis, in the small $\kappa$ regime, and chose parameters so that the second excited state energy was orders of magnitude larger. We then initialized the superposition $\ket{0_x} + \ket{1_x}$ (recalling that $H_0 = \kappa \sigma^x$ here), the lowest two eigenstates in the phase basis, and evolved time under a time varying charge offset $Q_{ext} = Q_0 \sin 2\pi \omega t$. We chose to simulate the very high frequency limit $\omega \geq 10 \kappa$; under such a drive, no transitions occur since we are far from any resonances, and the two qubit states evolve under the \emph{time averaged} Hamiltonian:
\begin{eqnarray}\label{Htavg}
H_{eff} = s \of{\barA} \sigma^x, \; s \of{\barA} \equiv \omega \int_{0}^{\omega^{-1}} dt' \cos \sqof{ \barA \sin \of{2 \pi \omega t'}}.
\end{eqnarray}
As shown in FIG, the Rabi flopping of the two states almost perfectly tracks the rate specified by Eq.~\ref{Htavg}, for a wide range of drive amplitudes. While the CEQ is driven resonantly at frequencies far below the scale where this effect comes into play for the closed system, it is an important check that the transverse field direction correctly tracks the oscillating charge offset in AC operation.

Finally, we note from this simulation that one could also achieve $1/f$ suppression in the CEQ by driving it at two or more very high frequencies $\omega_{1,2} \gg J$, where $\omega_{1}-\omega_{2} = \pm h$ and both tones are chosen to be far off resonant with any mixing with higher levels. In that regime, the calculation of ``optimal" base $\kappa$, drive amplitudes, and the AC-driven interaction with the bath all differ substantially from our predictions here. While we do not explore it here, this radically different operation regime could be an interesting topic for future research.

\subsection{Closed system simulations of Fermi golden rule (FGR) relaxation of CEQ}

We now turn to direct numerical simulations of CEQ relaxation. The main goal of this study is to establish the correspondence of matrix elements that underlies Eq. \ref{leerr}, and therefore ensure that we have quantiatively good estimates of coherence. Specifically, consider three distinct settings -- a single qubit coupled to a bath, a pair of strongly coupled spin-half degrees of freedom whose ground state doublet is the qubit, and then, finally, the latter driven resonantly, i.e. what we have referred to throughout this paper as the CEQ. Each of these three different physical systems is coupled to the bath via single spin terms. 

To isolate low energy relaxation processes in Eq. \ref{leerr} we initialize the bath in its ground state, hence making the groundstate of the qubit stable, while the excited state is expected to decay via an FGR mechanism, i.e. with a rate set by the product of the square of the relevant matrix element and the bath density of states (DOS). While this physics has been studied extensively for the first two cases \cite{gardinerzoller,albash2012quantum,yip2018quantum}, the latter case of the resonantly driven dimer is somewhat unusual, so we will now present a direct numerical analysis of the relaxation of such a transient Floquet problem\cite{ikeda2021fermi}.  While a physical bath is expected to have a special structure near its ground state, both in the density of states and matrix elements to excitations, we purposely design the simulation to lack such structure\cite{santra2017fermi,micklitz2022emergence}, so as to enable sharp focus on the matrix elements (physically, we are interested in finite temperature baths, of course). With that in mind, the bath Hamiltonian $H_B$ will be generated to have a flat spectrum and random eigenvectors. The latter are obtained by shuffling eigenvectors of random Hermitian matrices while the former is generated separately from a box distribution of width 10 (in units $J=1$).  Qubit couplings to the bath are engineered similarly, with the bath operators $M$ given by random Hermitian matrices.  We have simulated bath coupling to both of the individual spins, one of the three (x,y,z) channels at a time, while only driving one of them resonantly, i.e. at the frequency of the ground state doublet of the fully interacting dimer.  We only present the results of the y-coupling as this channel is expected to be the strongest practice (see main text) but, also, importantly, there aren't complications in interpreting the results (see below). Specifically, the Hamiltonian is taken to be 
\begin{align}
&H = H_{CEQ} +H_1 (t) + H_{B} + H_{int}, \\
&H_{int} = g_{y} (M_1 \sigma^y_1 + M_2 \sigma^y_2 )\\
&H_1(t) = \kappa [\cos (\alpha \sin\omega t) \sigma^x_1 +\sin (\alpha \sin\omega t) \sigma^y_1] 
\end{align}

\begin{figure}
    \centering
    \includegraphics[width=3in]{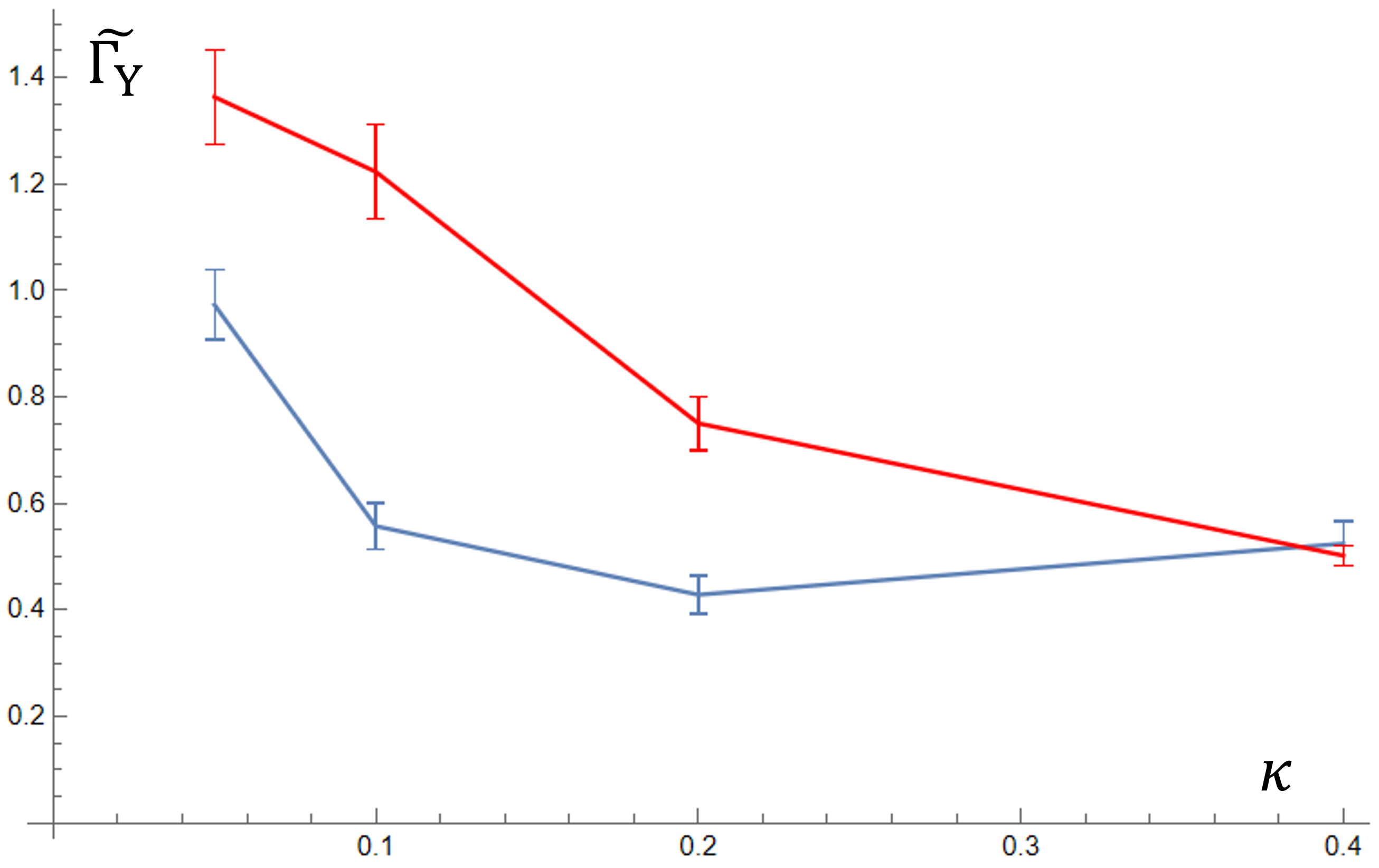}
    \caption{Twice the FGR constant is computed and plotted for the driven problem (red trace) vs. FGR constant for the undriven dimer (blue). See text for definitions.}
    \label{fig:fgrY}
\end{figure}

Decay rates are computed by observing and fitting linear in time decay of $P_1(t)\equiv \langle t| \of{ |1\rangle \langle 1| \otimes I_B } |t\rangle$, where $|t\rangle$ is the time evolved state of the combined system. Decay rates of the driven problem are computed stroboscopically, i.e. at integer Floquet periods of the $g=0$ problem. This approach works well at weak coupling and in the absence of diagonal drifts, i.e. if system bath coupling has a finite expectation value at short times and thus induces slow coherent dynamics, e.g. shifts of the apparent Floquet period. These latter complications are observed in coupling to $Z$ (and also $X$) terms, while the simple approach works well for the $Y$ coupling. To compare meaningfully we divide (normalize) the extracted decay rates by the product of $\delta^2$, the bath ground state expectation value of $m^2=[M^2]_{00}$ and the square of the matrix element of the q-bit flip operator, e.g. $|\langle 1|Y|0\rangle|^2$ -- we refer to such normalized decay rate as the FGR constant. Note, that while the bare decay rate is expected to vary strongly with parameters following $\delta^2 \kappa^L m^2$, i.e. several orders of magnitude for the range of $\delta$ and $\kappa$ we explored, the FGR constant plotted in Fig. 1 is nicely stationary around 1. Furthermore, we obtain clear evidence of the expected factor of 2 relationship between AC and DC relaxation. This is because the Floquet eigenstates of the resonantly driven CEQ are equal amplitude superpositions of ground-state and first excited states of the undriven logical qubit, hence the expected relaxation rate of these Floquet eigenstates is expected to be half of that of the undriven problem (since the ground state does not relax at zero temperature).

Interestingly, the observed DC error rates are somewhat smaller than expected in our perturbative analysis, with the AC error rates roughly equal. We attribute the AC error rate being larger (once the factor of 2 correction is applied) than the DC error rate to the likely presence of additional bath-assisted tunneling processes induced by the AC drive, as discussed in \cite{kapit2021noise}. In both cases the expected scaling with variation of CEQ parameters is roughly correct, so we are confident that our coherence estimates in the main text provide decent estimates of logical state coherence in a real experiment.

\end{document}